\documentclass[manuscript]{aastex}

\shorttitle{Star formation history of the Milky Way halo traced by the Oosterhoff dichotomy among globular clusters}
\shortauthors{Jang and Lee}

\begin{document}



\title{Star formation history of the Milky Way halo traced by the Oosterhoff dichotomy among globular clusters}

\author{Sohee Jang and Young-Wook Lee}
\affil{Center for Galaxy Evolution Research and Department of Astronomy, Yonsei University, Seoul 120-749, Korea }

\email{ywlee2@yonsei.ac.kr}





\begin{abstract}
In our recent investigation of the Oosterhoff dichotomy in the multiple population paradigm \citep{Jan14}, we have suggested that the RR Lyrae variables in the Oosterhoff groups I, II, and III globular clusters (GCs) are produced mostly by the first, second, and third generation stars (G1, G2, and G3), respectively. Here we show, for the first time, that the observed dichotomies in the inner and outer halo GCs can be naturally reproduced when these models are extended to all metallicity regimes, while maintaining reasonable agreements in the horizontal-branch type versus [Fe/H] correlations. In order to achieve this, however, specific star formation histories are required for the inner and outer halos. In the inner halo GCs, the star formation commenced and ceased earlier with relatively short formation timescale between the subpopulations ($\sim$0.5 Gyr), while in the outer halo, the formation of G1 was delayed by $\sim$0.8 Gyr with more extended timescale between G1 and G2 ($\sim$1.4 Gyr). This is consistent with the dual origin of the Milky Way halo. Despite the difference in detail, our models show that the Oosterhoff period groups observed in both outer and inner halo GCs are all manifestations of the ``population-shift'' effect within the instability strip, for which the origin can be traced back to the two or three discrete episodes of star formation in GCs.
\end{abstract}


\keywords{stars: horizontal-branch --- stars: variables: other --- Galaxy: formation --- globular clusters: general --- Galaxy: halo}



\section{Introduction}

\begin{figure}
\centering
\epsscale{1.1}
\plotone{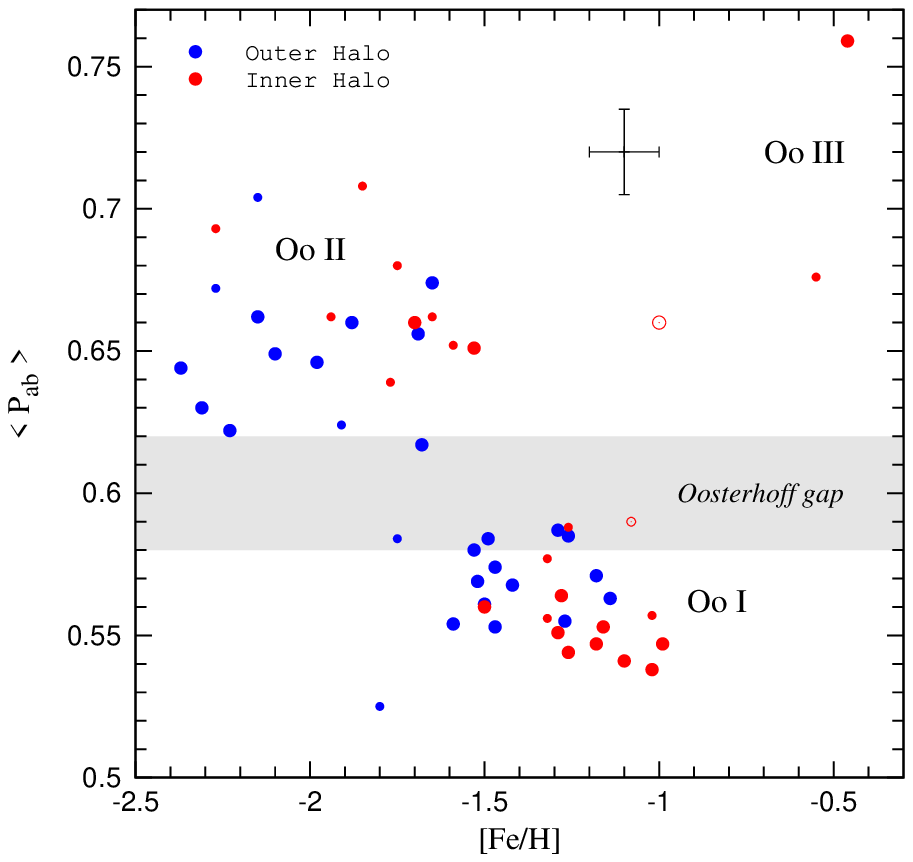}
\caption{The Oosterhoff period groups among GCs in the Milky Way. GCs are divided into three groups (Oo I, Oo II, and Oo III) according to $\left<P_{\rm ab}\right>$ and metallicity. The inner ($R_{gc}<$~8kpc) and outer ($R_{gc}\geq$~8kpc) halo GCs are shown as red and blue circles, respectively. The red open circles are two bulge GCs recently reported by \citet{Alo15}. The large (small) symbols are for clusters with a number of RR Lyrae stars $N_{ab}$ $\geqq$ 10 (3 $\leqq$ $N_{ab}$ $<$ 10). The typical error bar is shown and $\left<P_{\rm ab}\right>$ is in days. \label{fig1}}
\end{figure}

More than seven decades ago, \citet{Oos39} discovered that globular clusters (GCs) in the Milky Way are divided into two distinct groups according to the mean period ($\left<P_{\rm ab}\right>$)  of type ab RR Lyrae variables. The Oosterhoff group I GCs are relatively metal-rich, and have $\left<P_{\rm ab}\right>$ of $\sim$0.55 day, while more metal-poor group II GCs are observed to have longer $\left<P_{\rm ab}\right>$ of $\sim$0.65 day (see Figure~\ref{fig1}). Despite the importance of this phenomenon in our understanding of the Milky Way formation and Population II distance scale \citep[][and references therein]{San81,LDZ90,Yoo02,Smi04,Cat09}, this still remains one of the long-standing problems in modern astronomy. 

Numerous investigations were carried out to find the origin of the Oosterhoff dichotomy since its discovery. Among them, several deserve more detailed description. \citet{van73} suggested ``hysteresis mechanism", which explains the dichotomy as a difference in mean temperature between the type ab variables in the two groups. Sandage (1981), however, discovered the period-shift, the difference in period at given temperature, between the RR Lyrae stars in the two groups, which led him to conclude that the dichotomy is caused by the luminosity difference. By employing the synthetic horizontal-branch (HB) models, \citet{LDZ90} found that RR Lyrae variables highly evolved from the zero-age horizontal-branch (ZAHB) can reproduce the period-shift for the group II GCs if HB morphology is sufficiently blue. While this evolution effect can play an important role in the Oosterhoff dichotomy \citep[see][]{LeeZinn90}, it is not clear whether this effect alone could reproduce the observed dichotomy, because HB types of some group II GCs are too red to have enough evolution effect.

All of these investigations were performed on an assumption that GCs host only single population \citep[see also][]{Sol14}. Recent observations and population models, however, have convincingly demonstrated the presence of multiple populations in most GCs \citep[][and references therein]{Lee99,Bed04,Dan04,Lee05,Pio07,Car09,Han09,JW09,Mar09,Gra12a}. This has motivated \citet{Jan14} to suggest that the origin of the Oosterhoff dichotomy is intimately related to the presence of multiple populations in GCs. According to this new paradigm, the instability strip (IS) in the metal-poor group II GCs is populated by second-generation stars (G2) with enhanced helium and CNO abundances, while the RR Lyrae stars in the relatively metal-rich group I GCs are produced mostly by first-generation stars (G1) without these enhancements. This ``population-shift" within the IS with increasing metallicity can create the observed period-shift between the two groups. Synthetic HB models presented by \citet{Jan14}  were focused on the two outer halo GCs representing two Oosterhoff groups, M15 and M3. It was, therefore, not clear whether this population-shift effect applies to all GCs in addition to these two GCs, and also whether there is any difference between the inner and outer halo GCs. The purpose of this paper is to extend our multiple population models to all metallicity regimes to see whether the population-shift effect can reproduce the observed period dichotomy, and to suggest, based on $\left<P_{\rm ab}\right>$ versus [Fe/H] correlations, new constraints on the star formation time scales for the inner and outer halo GCs. 

\section{Data Collection and Model Construction}

 Table~\ref{tab1} lists Milky Way GCs containing 3 or more type ab RR Lyrae variables. Metallicity ([Fe/H]) and galactocentric distance ($R_{\rm gc}$, in kpc) are from 2010 edition of \citet{Har96} catalogue. Data for the mean periods (in days) and numbers of RR Lyrae stars ($\left<P_{\rm ab}\right>$, $\left<P_{\rm c}\right>$, $N_{ab}$, and $N_{c}$) are from the sources listed in the table, where double mode RR Lyrae variables, if any, are treated as c type variables \citep[see][]{Cle01}. For the GCs without the source listed in the table, they are from the compilations by \citet{Cat09} and \citet{Cle01}. Updated data for the HB type \citep{LDZ94} are mostly from H. Gim \& Y.-W. Lee (2009, private communication) and from our own compilation.
 
Figure~\ref{fig1} shows the Oosterhoff-Arp diagram based on the data in Table~\ref{tab1}. As is clear from this figure, Oosterhoff group II GCs are relatively metal-poor, while group I GCs are more metal-rich. However, two most metal-rich GCs (NGC 6388 and NGC 6441) have longest $\left<P_{\rm ab}\right>$ and are classified as group III \citep{Pri03}. The red open circles are 2Mass-GC02 and Terzan 10, two bulge GCs recently reported by \citet{Alo15}. Note that there is an abrupt increase in $\left<P_{\rm ab}\right>$ at [Fe/H] $\approx$ -1.5 dex, and that the inner halo GCs are in the mean more metal-rich than those in the outer halo.
 
In order to reproduce the observed period dichotomy, we have constructed a series of synthetic HB models with RR Lyrae variables, following the technics developed by \citet{LDZ90,LDZ94}, and as updated by \citet{Joo13} and \citet{Jan14} to include the effects of multiple populations. Our models are based on the most updated Yonsei-Yale ($Y^2$) isochrones and HB evolutionary tracks with enhanced helium and CNO abundances \citep{Han09,Lee15}, all constructed under the assumption that [$\alpha$/Fe] = 0.3 dex. These isochrones and evolutionary tracks are available from the corresponding author upon request. \citet{Rei77} mass-loss parameter $\eta$ was adopted to be 0.50, which best reproduced observed HB morphology versus [Fe/H] relation of the inner halo GCs when the mean age was assumed to be 13 Gyr \citep[see][for a similar result]{McD15}.\footnote{The value of $\eta$ depends on the adopted absolute age for the GCs. If we had adopted $\eta$ = 0.42 \citep{Jan14}, the mean age for the inner (old) halo GCs required to fit the observed HB morphologies would be $\sim$ 14 Gyr.} Following \citet{Jan14}, mass dispersion on the HB was adopted to be $\sigma_{M}$ = 0.010 $M_{\sun}$ for each subpopulation. 

For the RR Lyrae variables, the first overtone blue edge of the instability strip was taken from \citet{Tug72} with corrections for convection \citep{Ste84,LDZ90}, which agrees, to within the uncertainty in the reddening ($\Delta$(B - V) $\approx$ 0.02), with more recent calculations by \citet{Bon95}. As for the transition temperature between ab and c type variables, \citet{Bon95} showed that the transition occurs close to the first overtone red edge in M15 (group II), whereas it occurs near the fundamental blue edge in M3 (group I), which suggest that the hysteresis mechanism \citep{van73} may play an important role in the ``either-or" zone of the IS. Even if this theoretical scenario is able to reproduce many observational behaviors \citep{Bon97,Mar03}, some residual uncertainties on the first overtone red edge due to the significant dependence on the convection treatment may remain. An alternative approximated approach to this problem, adopted in our calculation, is to use the \citet{San06}'s semi-empirical equation (his eq. 17) for the blue edge of type ab variables as a transition temperature between the two types of RR Lyrae stars. It appears that some hysteresis effect is empirically reflected in his equation, as we obtain cooler temperatures ($\Delta log~T_{eff} \approx$ 0.008) for more metal-poor group II GCs from his equation.
The width of the instability strip was taken to be $\Delta log~T_{eff}$~=~ 0.060 \citep{Roo73,Ste84}, which best reproduces the observed value of $\left<P_{\rm ab}\right>$ in our models, and is acceptable to within the uncertainty of the observed width in M3 and M15 \citep[see Figures 2 \& 3 of][]{Bon95}. The absolute value of $\left<P_{\rm ab}\right>$ would be increased by $\sim$ 0.02 day if we had adopted $\Delta log~T_{eff}$~=~0.070, but this has only negligible effect on the relative difference in $\left<P_{\rm ab}\right>$ between the two Oosterhoff groups. The periods of fundamental mode RR Lyrae stars $\left(P_{\rm ab}\right)$ were calculated from \citet{van71} and those of first overtone mode variables $\left(P_{\rm c}\right)$ were obtained from the equation by \citet{Bon97}. In order to minimize the stochastic effect stemming from a finite number of stars, we obtained mean values of 1000 simulations, each with the sample size of 10000 HB stars, at given parameter combination.
 
\begin{figure}
\centering
\epsscale{0.8}
\plotone{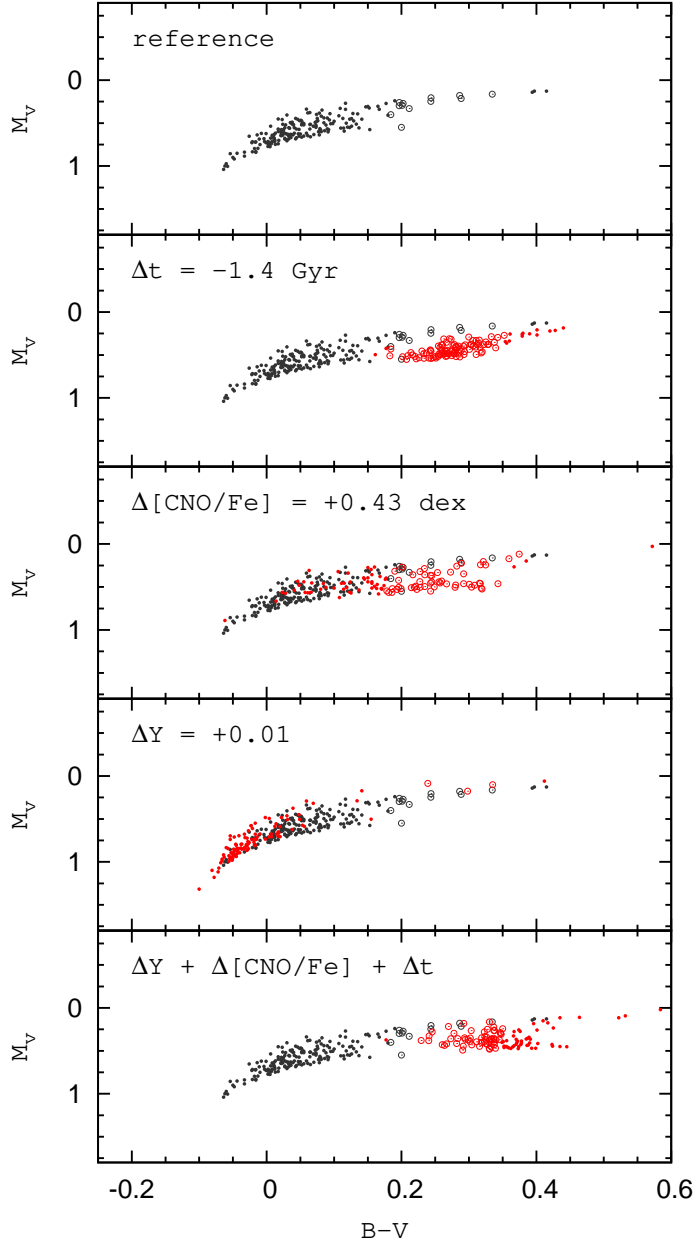}
\caption{The effects of age, CNO and helium abundances on the HB morphology. Compared to the reference case ([Fe/H] = -1.9 dex, age = 12.4 Gyr, $Y$ = 0.23), which is repeated in each panel with dark-gray circles, the effects of age, [CNO/Fe], and helium abundance are illustrated respectively with red circles. The bottom panel shows the net effect of these three parameters operating simultaneously, which confirms that HB morphology of G2 can be redder than that of G1 when the helium enhancement is mild. Open circles are RR Lyrae variables.\label{fig2}}
\end{figure}

The major assumption made in \citet{Jan14} was the presence of G2, which is mildly enhanced in helium and CNO abundances. In terms of helium enhancement, this subpopulation is in between the G1 (with normal helium abundance) and super-He-rich third generation stars (G3, if any). Both theories and observations suggest that these G2 and G3 would be enhanced in both helium and CNO abundances from the chemical pollution and enrichment by intermediate-mass asymptotic-giant-branch stars and/or fast-rotating massive-stars \citep{Fen04,Coh05,Dec07,Dec09,Ven09,Yon09,Yon15,Kar12,Alv12,Gra12a,Mar12}. In general, helium enhanced subpopulations, like G3 in our models, are placed on the bluer HB \citep{LDZ94,Dan04,Lee05,Gra11,Gra12b,Vil12,Joo13,Kun13a,Mar14}. However, when the helium enhancement is relatively small ($\Delta$$Y$ $\approx$ 0.01), as is required for the RR Lyrae variables in M15 \citep{Jan14}, G2 can be placed on the relatively redder HB, because the age and CNO effects can overwhelm the helium effect. This is illustrated in Figure~\ref{fig2}, where the effects of age, [CNO/Fe], and helium abundance on the HB morphology are presented using our synthetic HB models. First, the HB morphology is getting redder as age decreases because of the higher mean mass of HB stars at younger age. Similarly, the morphology of HB gets redder as CNO abundance increases, since the efficiency of hydrogen shell burning and opacity increase with increasing [CNO/Fe]. However, as helium abundance increases, the HB morphology gets bluer, mainly because helium-rich stars evolve more rapidly, and therefore, have lower masses at the tip of the red giant-branch (RGB) and HB for a given age. The bottom panel shows the net effect of these three parameters operating simultaneously. This confirms that the HB morphology of G2 can be redder than that of G1 when the helium enhancement is modest (see also Figure~\ref{fig7} below). The effects of CNO and helium abundances on the periods of RR Lyrae stars are presented in Figure 3 of \citet{Jan14}.

\section{Comparison of Models with Observations}

In this section, we extend our models to all metallicity regimes, covering the inner and outer halo GCs, to see whether the population-shift effect can reproduce the observed period dichotomy. Our previous modeling \citep{Joo13,Jan14} shows that the relative placement of subpopulations (G1, G2 and G3) on the HB is qualitatively similar in most GCs, but parameter combinations (age, $Y$, and [CNO/Fe]) defining these subpopulations can be somewhat different from one GC to another. Therefore, here, it is our goal to find the representative parameter combinations for the inner and outer halo GCs respectively, which can best reproduce the observed $\left<P_{\rm ab}\right>$ versus [Fe/H] correlations in those regimes, while maintaining reasonable agreements in the observed HB type versus [Fe/H] relations.

First, we constructed, with a parameter combination found by \citet{Jan14} from an outer halo GC M15, a series of synthetic HB models by varying [Fe/H] from -2.3 to -0.8 dex. Preliminary comparison of this model sequence with the outer halo GCs on the $\left<P_{\rm ab}\right>$ versus [Fe/H] plane (see Figure~\ref{fig3}) indicated that the presence of two distinct $\left<P_{\rm ab}\right>$ groups can be reproduced qualitatively, but some fine tuning of input parameters was required for the best-fit with the data. A solid line in Figure~\ref{fig3} is our best-fit models for the outer halo, and Table~\ref{tab2} lists the best-fit input parameters obtained from this endeavour (see Figure~\ref{fig6} below for the corresponding models in the HB type versus [Fe/H] plane).\footnote{When log Z was converted to [Fe/H] in our modeling, we provisionally adopted [$\alpha$/Fe] = 0.0 dex, because otherwise the sudden change in $\left<P_{\rm ab}\right>$ occurred at [Fe/H] that is somewhat metal-poor compared to the observed trend, and the model locus was not well matched with the data at the metal-rich regime. Of course, this is not consistent with the observed [$\alpha$/Fe] = 0.2 - 0.3 dex \citep{Pri05}, and therefore is equivalent to the shift of [Fe/H] by 0.14 - 0.2 dex, which, however, is within the current uncertainty in the metallicity scale. This shift is required because our $\alpha$, CNO, and He enhanced HB tracks have excessively extended blueward loops at given [Fe/H]. We suspect that this is due to the implicit assumption of fully efficient semi-convection adopted in our HB stellar models, which was introduced in 70s to match the observed color spread on the HB and the primordial helium abundance \citep[see, e.g.,][]{Swe72}, well before the discovery of He enhanced multiple populations in GCs. } Note that $\Delta$$Y$ and $\Delta Z_{CNO}$ between G1 and G2 mostly determine the values of $\left<P_{\rm ab}\right>$ for the two Oosterhoff groups \citep[see Figure 3 of][]{Jan14}, while $\Delta$t(G1 - G2) mainly controls the overall feature of the distribution in $\left<P_{\rm ab}\right>$ versus [Fe/H] diagram. Our best-fit models were constructed under the assumption that the age difference between G1 and G2, $\Delta$t(G1 - G2), is 1.4 Gyr. We found that this age difference provides the best match with the observed GCs in the outer halo, both for the groups I and II. When the age difference is different from this value, for example, $\Delta$t(G1 - G2) = 0.5 Gyr as shown by the dashed line, the models fail to reproduce the observed distribution, which illustrates the sensitivity of our models to $\Delta$t(G1 - G2). Similarly, if the separation between G1 and G2 on the HB was obtained by increasing $\Delta Z_{CNO}$ instead of $\Delta$t(G1 - G2), we would obtain $\Delta$$\left<P_{\rm ab}\right>$ that is too large (by $\sim$ 0.06 day) compared to the observations. Our best fit models (the solid line) show a ``wavy" feature in the group II regime (-2.3 $\la$ [Fe/H] $\la$ -1.8 dex), producing GCs having $\left<P_{\rm ab}\right>$ of $\sim$0.65 day along this locus. Inspection of synthetic HB models indicates that this feature is dictated by the placement of G2 within the IS and the contribution of evolved stars from either side of the IS.

We have then applied this parameter combination determined from the outer halo to the inner halo GCs to see whether this can be similarly employed to all GCs. The HB types for the inner halo GCs are systematically bluer than those for the outer halo GCs (see Figure~\ref{fig6} below), and the general consensus is that this global second parameter phenomenon stems from a small age difference between the inner and outer halo GCs \citep{LDZ94,Dot10,Gra10}. Therefore, in order to reproduce the HB types of the inner halo GCs, first, we have adopted $\Delta$t(G1 - G2) to be 1.4 Gyr as was found from the outer halo GCs, but constructed models under the condition that the inner halo GCs are $\sim$1 Gyr older than those in the outer halo. Other parameters, $\Delta$$Y$, $\Delta$$Z_{CNO}$, and mass-loss parameter $\eta$, are held identical to the values adopted for the outer halo. In this case, however, we fail to reproduce the observed distribution of $\left<P_{\rm ab}\right>$ among the inner halo GCs (see the dashed line in Figure~\ref{fig4}). Compared to the observations, the jump in $\left<P_{\rm ab}\right>$ occurs at more metal-rich regime ([Fe/H] $\approx$ -1 dex), and the ``plateau" of $\left<P_{\rm ab}\right>$ distribution is formed at too low value of $\left<P_{\rm ab}\right>$. Therefore, by constructing models under different assumptions on the value of $\Delta$t(G1 - G2), we found that models can best match the observations when $\Delta$t(G1 - G2) = 0.5 Gyr (see the solid line in Figure~\ref{fig4}). The observed plateau of $\left<P_{\rm ab}\right>$ in -1.8 $\la$ [Fe/H] $\la$ -1.6 dex is also naturally reproduced in this case, which is explained in detail in Figures~\ref{fig9} and \ref{fig10} below. Our best-fit input parameters for the inner halo GCs are listed and compared with those for the outer halo in Table~\ref{tab2}, and Figure~\ref{fig5} compares our best-fit models with the observations both for the inner and outer halo GCs. 

Our models were constructed to reproduce not only the observed mean periods of RR Lyrae stars, but also the HB types of halo GCs, as presented in Figure~\ref{fig6} and Table~\ref{tab3}. For example, our best-fit models in Figure~\ref{fig5} were constructed along the solid lines (G1+G2 models) in Figure~\ref{fig6}. Unlike $\left<P_{\rm ab}\right>$, the HB type is sensitively affected by the presence of G3 in all metallicity regimes, and therefore, models including G3 are also presented in Figure~\ref{fig6} (see the dashed lines). As is clear from the figure, most of the observed GCs are reasonably well reproduced by our ``G1+G2" or ``G1+G2+G3" models. Note that some of the outer halo GCs show inner halo (``old halo") characteristics in terms of HB morphology. In order to illustrate how G1, G2 and G3 would affect the final morphology of HB, we present in Figure~\ref{fig7} the synthetic HB models at three different metallicities along the model lines in Figure~\ref{fig6}. In the outer halo, where the separation between G1 and G2 is relatively large, GCs undergo the population-shift more clearly as a subpopulation occupying the IS is progressively changing with metallicity \citep[see also Figure 1 of][]{Jan14}. In the inner halo GCs, however, we can see that G1 and G2 are more overlapped in the IS because the separation between G1 and G2 is small. For the GCs with G3, HB stars belonging to this additional subpopulation are represented as filled gray circles in Figure~\ref{fig7} by employing the parameters listed in Table~\ref{tab2}. Note, however, that helium abundances of G3 in real GCs would be somewhat different from one GC to another.

As described above, and shown in Figure~\ref{fig8}, the most obvious features in our models for the inner halo GCs are (1) the sudden jump of $\left<P_{\rm ab}\right>$ at [Fe/H] $\approx$ -1.4 dex, (2) the plateau of $\left<P_{\rm ab}\right>$ distribution in the metallicity range of -1.8 $\la$ [Fe/H] $\la$ -1.6 dex, and (3) the second jump of $\left<P_{\rm ab}\right>$ at [Fe/H] $\approx$ -1.8 dex. We believe that this ``two-stage jump" in $\left<P_{\rm ab}\right>$ is the most important effect we found for the complete understanding of the Oosterhoff dichotomy in the inner halo GCs. As compared in Figure~\ref{fig8}, the first jump in $\left<P_{\rm ab}\right>$ at [Fe/H] $\approx$ -1.4 dex \citep[as first discovered by][]{LeeZinn90,Yoo02} is well reproduced by both ``G1 only" and ``G1+G2" models, while the plateau in Oosterhoff II regime is only reproduced in the multiple population paradigm (G1+G2 model).  In order to explain this effect in more detail, we illustrate, in Figure~\ref{fig9}, four synthetic HB models with evolutionary tracks in the relevant metallicity range of -2.0 $\la$ [Fe/H] $\la$ -1.4 dex.  At [Fe/H] $\ga$ -1.4 dex, type ab RR Lyrae zone is mostly occupied by HB stars on or near ZAHB from both G1 and G2, producing low values of mean luminosity and $\left<P_{\rm ab}\right>$. As metallicity decreases, HB morphology is progressively getting bluer, and the ZAHB portion of G1 is crossing the fundamental blue edge (FBE), leaving the zone mostly occupied by He and CNO enhanced G2, which increases $\left<P_{\rm ab}\right>$ to $\sim$0.65 day (see the second panel). This state continues with little change in $\left<P_{\rm ab}\right>$ until [Fe/H] $\approx$ -1.8 dex.  With a further decrease in metallicity, ZAHB portion of G2 is also getting out of the zone, and therefore only highly evolved stars from the blue side of the FBE penetrate back into the zone, increasing again mean luminosity and $\left<P_{\rm ab}\right>$. Figure~\ref{fig10} shows a schematic diagram that explains these sequential departures of G1 and G2 ZAHBs toward higher $T_{eff}$ with decreasing metallicity, which lead to the two-stage jump in $\left<P_{\rm ab}\right>$. Note that this effect is another manifestation of the population-shift effect \citep{Jan14} in the inner halo, where, unlike the outer halo, the separation between G1 and G2 is small on the HB. Note further that our models for the inner halo GCs correctly predict that the population shift, and therefore the jump in $\left<P_{\rm ab}\right>$, occurs at bluer HB type compared to the outer halo GCs (see also Figures~\ref{fig6} and \ref{fig7}). 

As suggested by previous investigators \citep{Cal07,Yoo08,Jan14}, for the case of more metal-rich Oosterhoff group III GCs such as NGC 6441 and 6388, the HB stars cannot be populated within the IS unless helium abundance is significantly enhanced. Therefore, we present our models including more helium-rich G3 in Figure~\ref{fig11}, and the best-fit parameters adopted for G3 are also listed in Table~\ref{tab2}. Our models with G3 are qualitatively similar to those in \citet{Jan14}, but here we include the spread in helium abundance in order to reproduce the observations more realistically. As metallicity increases, G1 and G2 cross the red edge of the IS, mostly settling down in the red HB, while helium-rich G3 penetrate into the IS from blue HB, producing RR Lyrae stars with longer periods. As shown in Figure~\ref{fig11}, the presence of G3 starts to have influences on $\left<P_{\rm ab}\right>$ from [Fe/H] $\approx$ -1.5 dex for the outer halo, while they have effects from [Fe/H] $\approx$ -1 dex for the inner halo. With a further increase in metallicity, more and more helium enhanced G3 are progressively penetrating into the IS, continuously increasing $\left<P_{\rm ab}\right>$. As is evident from Figure~\ref{fig11}, group III GCs and some group I GCs in the outer halo are better reproduced by our models with G3. Note, however, that our models with G3 were calibrated with only three GCs (NGC 6441, 6388, and Terzan 10) for which type ab variables are predicted to be fully produced by G3, and therefore, parameters for G3 listed in Table~\ref{tab2} are more uncertain compared to those for G1 and G2.



\section{Discussion}

We have shown that a possible explanation of the observed Oosterhoff dichotomy can be naturally obtained when the multiple population models of \citet{Jan14} are extended to all metallicity regimes covering the inner and outer halo GCs. We found that the population-shift responsible for the Oosterhoff dichotomy can be traced back to the two or three discrete episodes of star formation in GCs. Interestingly, in order to reproduce the observed period dichotomies for the inner and outer halo GCs, respectively, different star formation timescales are required. The best-fit models suggest that the star formation in the inner halo GCs commenced and completed earlier within the timescale of $\sim$0.5 Gyr, while that for the outer halo GCs was delayed by $\sim$0.8 Gyr with more extended timescale ($\sim$1.4 Gyr) between G1 and G2. Despite the difference in detail, our models show that the Oosterhoff period groups observed in the inner and outer halo GCs are all manifestations of the population-shift effect within the instability strip.

Further supporting evidence for our models is provided by the observed difference in mean periods ($\left<P_{\rm ab}\right>$ - $\left<P_{\rm c}\right>$) between type ab and type c variables (see Figure~\ref{fig12}). Interestingly, the Oosterhoff group II GCs are divided into two subgroups in this diagram, which is not obvious in $\left<P_{\rm ab}\right>$ (see Figure~\ref{fig11}). Following \citet{Yoo02}, we call these subgroups ``Oo IIa" for GCs with [Fe/H] $\ga$ -1.9 dex and ``Oo IIb" for those with [Fe/H] $\la$ -1.9 dex. Our models can explain this rather clearly. The c type variables in Oo IIa subgroup are populated by both G1 and G2, and thus have relatively low $\left<P_{\rm c}\right>$, while those in Oo IIb are mostly composed of G2 which leads to relatively higher value of $\left<P_{\rm c}\right>$ and smaller difference in $\left<P_{\rm ab}\right>$ - $\left<P_{\rm c}\right>$. In addition to $\left<P_{\rm ab}\right>$, the fraction of c type variables ($f_{c}$) is, in general, also different between the two Oosterhoff groups. Although our models predict $f_{c} \approx$ 0.4 - 0.5 for the group II GCs and $f_{c}\approx$ 0.2 for more metal-rich group I GCs, which apparently agree with the observations, this result is subject to the uncertainties in the adopted blue edges, transition temperature between the two types of variables, and the width of the IS. Inspection of our models indicates that the difference in $f_{c}$ between the two groups is mostly due to the uneven distribution in $T_{eff}$ of RR Lyrae stars within the IS, which is qualitatively consistent with the conclusion from single population models by \citet{LDZ90}.

Our models predict that there is an age difference of $\sim$1 Gyr on average between the inner and outer halo GCs, excluding some outer halo GCs with predominantly blue HB type \citep[those with ``old halo" characteristics as defined by][]{Zin93} which would be as old as the inner halo GCs. This is consistent with the main-sequence age dating based on the recent HST data \citep{Dot10}, which suggests an age difference of $\sim$0.8 Gyr between the corresponding samples of GCs. Note, however, that ages obtained from such studies do not yet explicitly account for the effects of multiple populations in GCs, and therefore, variations in He and CNO abundances among subpopulations should be taken into account for better age determinations of GCs. Direct age dating of G1 and G2 in a given GC would be even more challenging, because, as illustrated in Figure 3 of \citet{Jan14}, the effect of younger age for G2 would be cancelled out by the effect of enhanced CNO on the isochrones. Considering this, perhaps the most important result predicted by our models from the Oosterhoff dichotomy is that the star formation timescale between G1 and G2 is more extended ($\sim$1.4 Gyr) in the outer halo GCs compared to that of the inner halo ($\sim$0.5 Gyr). This result for the outer halo GCs is qualitatively consistent with a more prolonged star formation history observed in the Local Group dwarf galaxies \citep[see, e.g.,][]{Tol09}. This is most likely because of the suggested difference in the formation history, in which the inner halo was formed mostly by dissipational merger and collapse, while dissipationless accretion of stars from primordial building blocks and/or satellite galaxies were more important in the outer halo \citep{Zin93,Bek01,Lee07,Car07,Zol09}. In this regard, it is important to note that \citet{Yoo02} found a planar alignment of Oo IIb GCs, which suggests a captured origin from a satellite galaxy. \citet{Paw12} went further from this and suggested that most of the outer halo GCs would belong to the Vast Polar Structure of satellite galaxies.

According to our results, the type ab variables in the Oosterhoff group II GCs are produced mostly by G2, while those in group I GCs are more populated by G1. The fact that the RR Lyrae variables with Oosterhoff I and II characteristics are similarly present in the halo field \citep{Lee90,San90,Sun91,Cat04,Ses13} would therefore suggest that building blocks similar to GCs (in terms of stellar populations) provided the halo field with RR Lyrae stars originated from both G1 and G2. Unlike the super-helium-rich G3 that require very extreme formation conditions in the central region of a proto-GC \citep[see, e.g.,][]{Der08}, the G2 in our models experienced just a modest helium enhancement ($\Delta$$Y$ $\approx$ 0.01), and therefore, they could have been formed ubiquitously in a proto-GC. Thus, it appears that both G1 and G2 were effectively provided to the halo field when merging and disruption of building blocks were much more active in the early phase of the Milky Way formation.

\begin{figure}
\epsscale{1.0}
\plotone{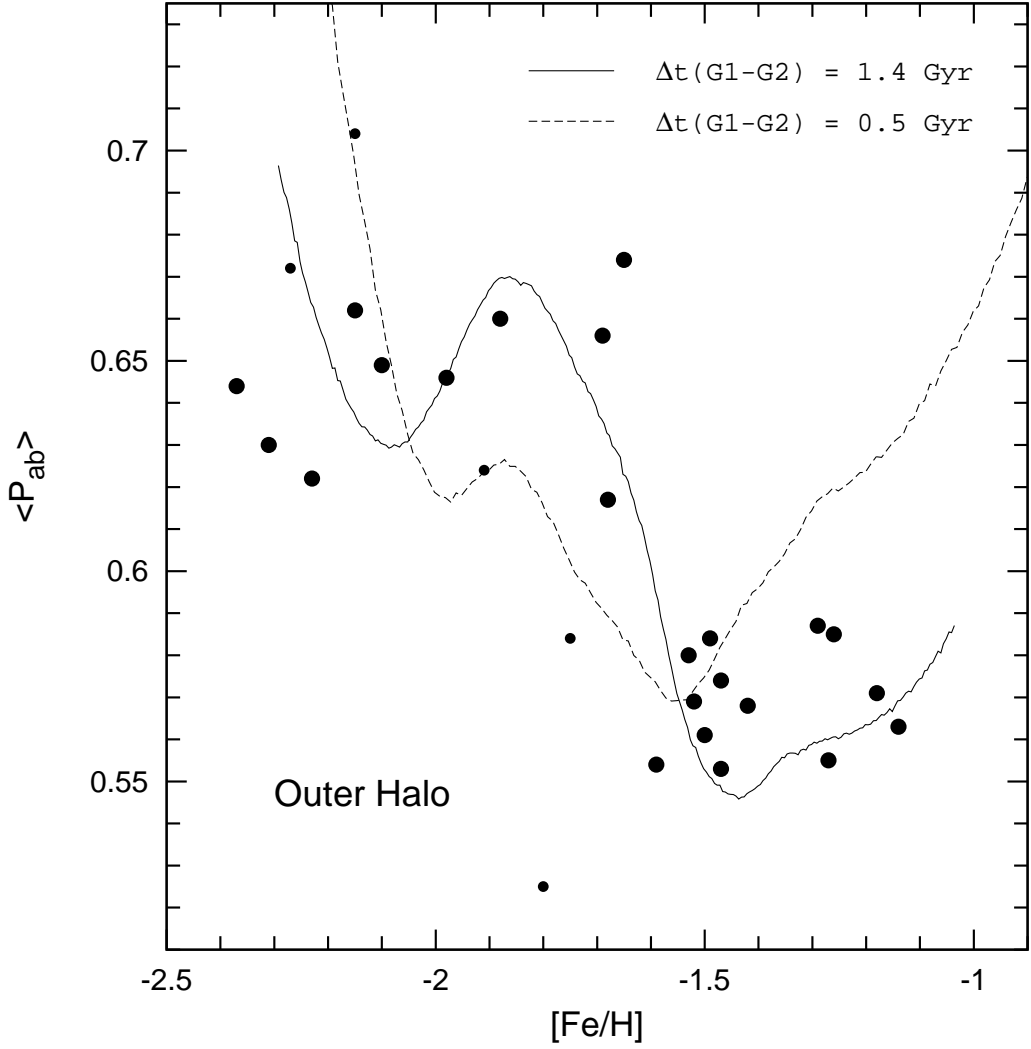}
\caption{Comparison of the outer halo GCs with our models in the $\left<P_{\rm ab}\right>$ versus [Fe/H] diagram.  The solid and dashed lines are our models constructed under different assumptions on the age difference between G1 and G2, $\Delta$t(G1 - G2). The best match is obtained when $\Delta$t(G1 - G2) = 1.4 Gyr (see the text). Definitions for the large and small symbols are as in Figure~\ref{fig1} and $\left<P_{\rm ab}\right>$ is in days.}\label{fig3}
\end{figure}

\begin{figure}
\epsscale{1.0}
\plotone{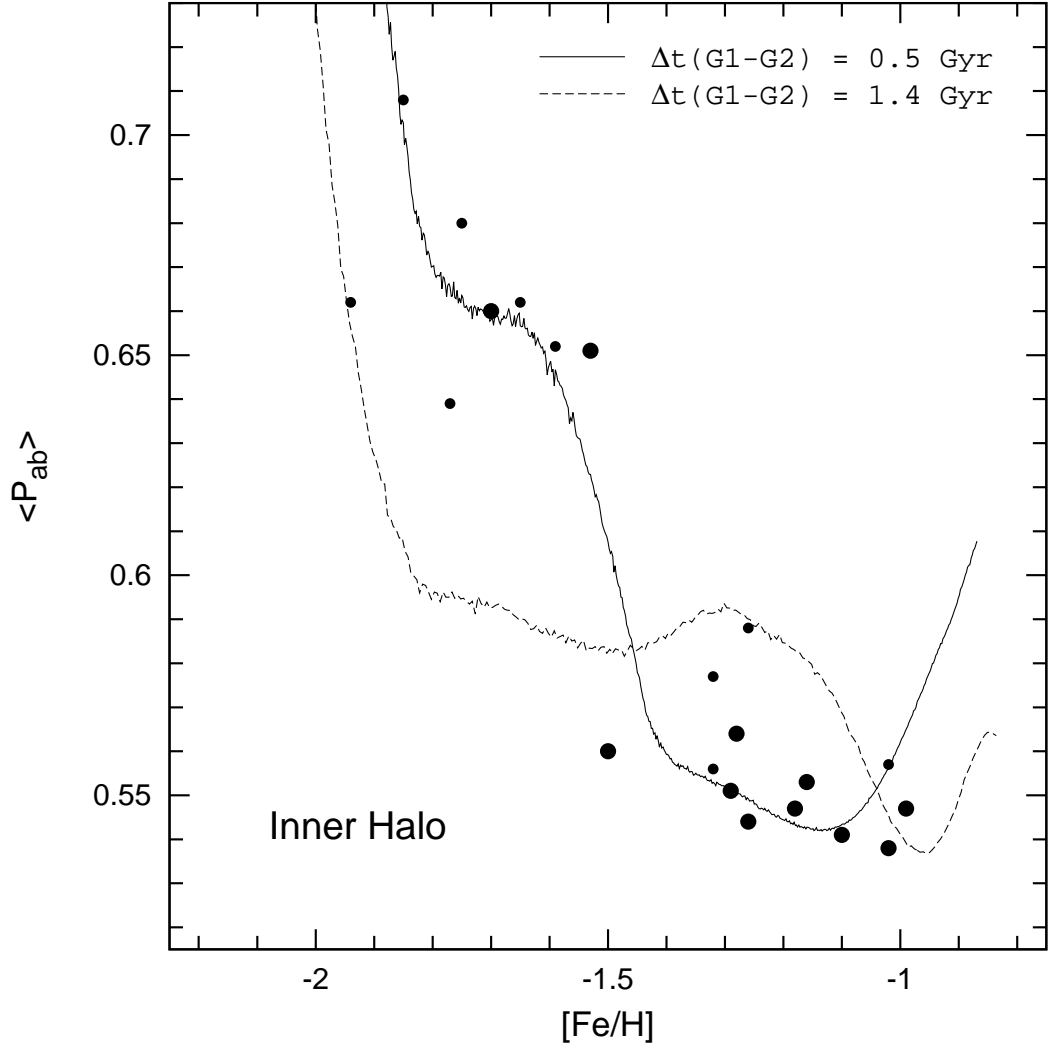}
\caption{Same as Figure~\ref{fig3}, but for the inner halo GCs. The best match is obtained when $\Delta$t(G1 - G2) = 0.5 Gyr.\label{fig4}}
\end{figure}

\begin{figure}
\epsscale{1.05}
\plotone{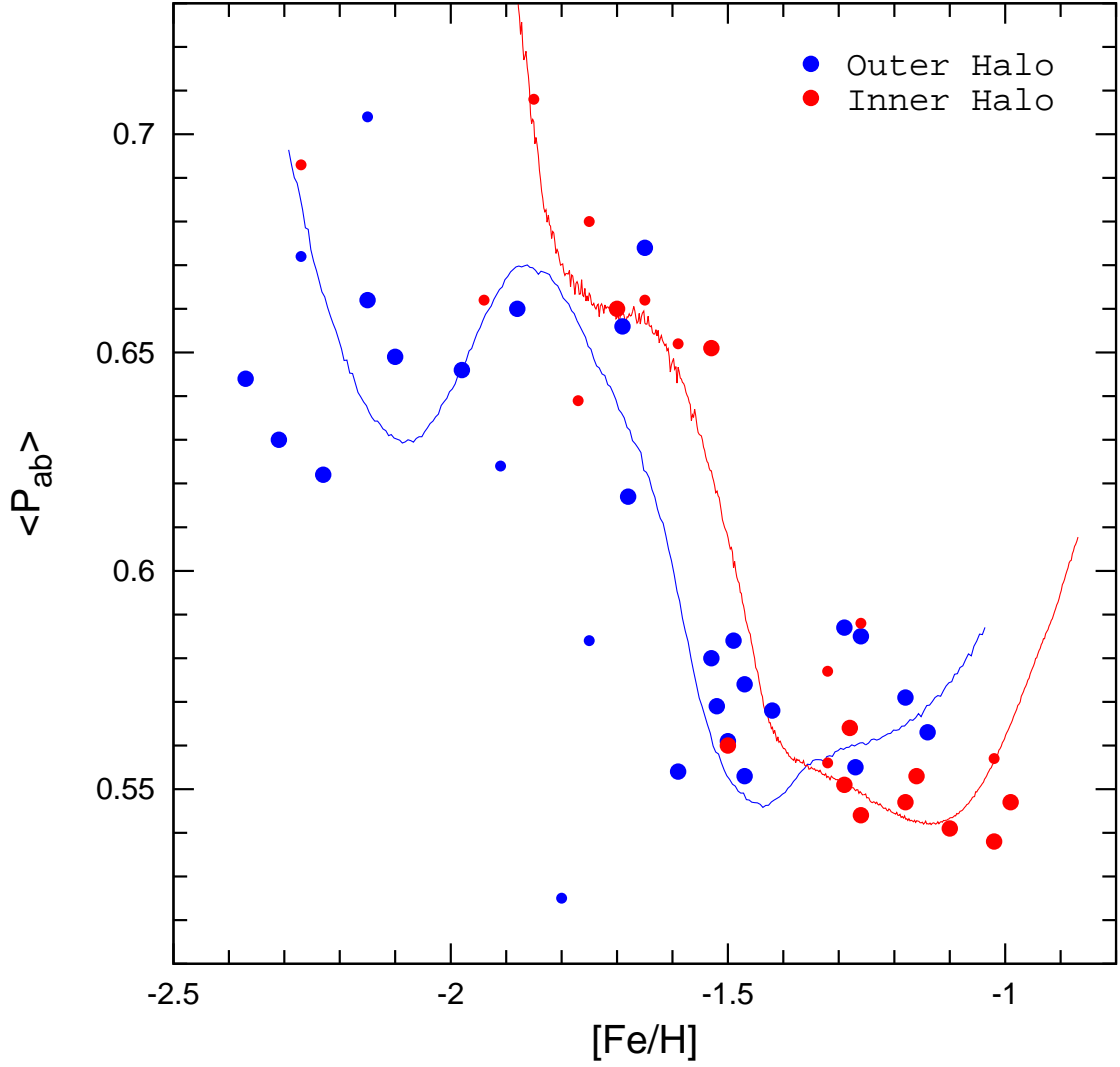}
\caption{Same as Figures~\ref{fig3} and \ref{fig4}, but only best-fit models (with G1 and G2) for the inner and outer halo GCs are compared with the observations. \label{fig5}}
\end{figure}

\begin{figure}
\epsscale{1.05}
\plotone{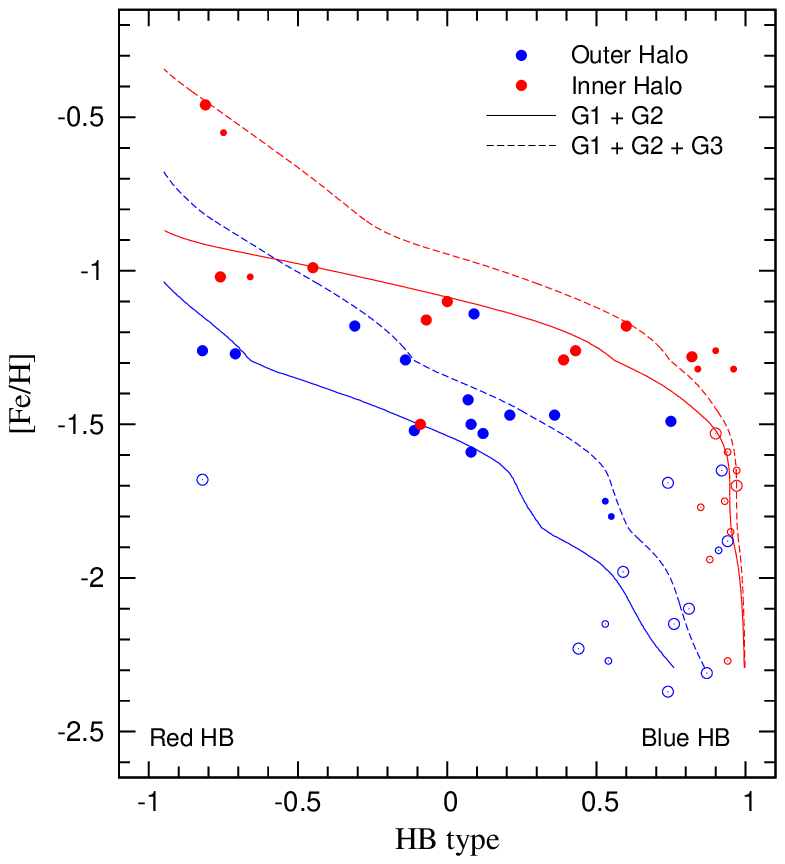}
\caption{The HB morphology versus metallicity diagram \citep{LDZ94} for our GC sample. Our best-fit models for the inner ($R_{gc}<$~8kpc) and outer ($R_{gc}\geq$~8kpc) halo GCs are shown as red and blue lines. The solid and dashed lines are for ``G1+G2" and for ``G1+G2+G3", respectively (see Table~\ref{tab2} for the adopted parameters). The corresponding models in the $\left<P_{\rm ab}\right>$ versus [Fe/H] plane are in Figures~\ref{fig5} and \ref{fig11}. Definitions for the large and small symbols are as in Figure~\ref{fig1}, but open circles are for the Oosterhoff group II GCs.\label{fig6}}
\end{figure}

\begin{figure}
\epsscale{1.23}
\plotone{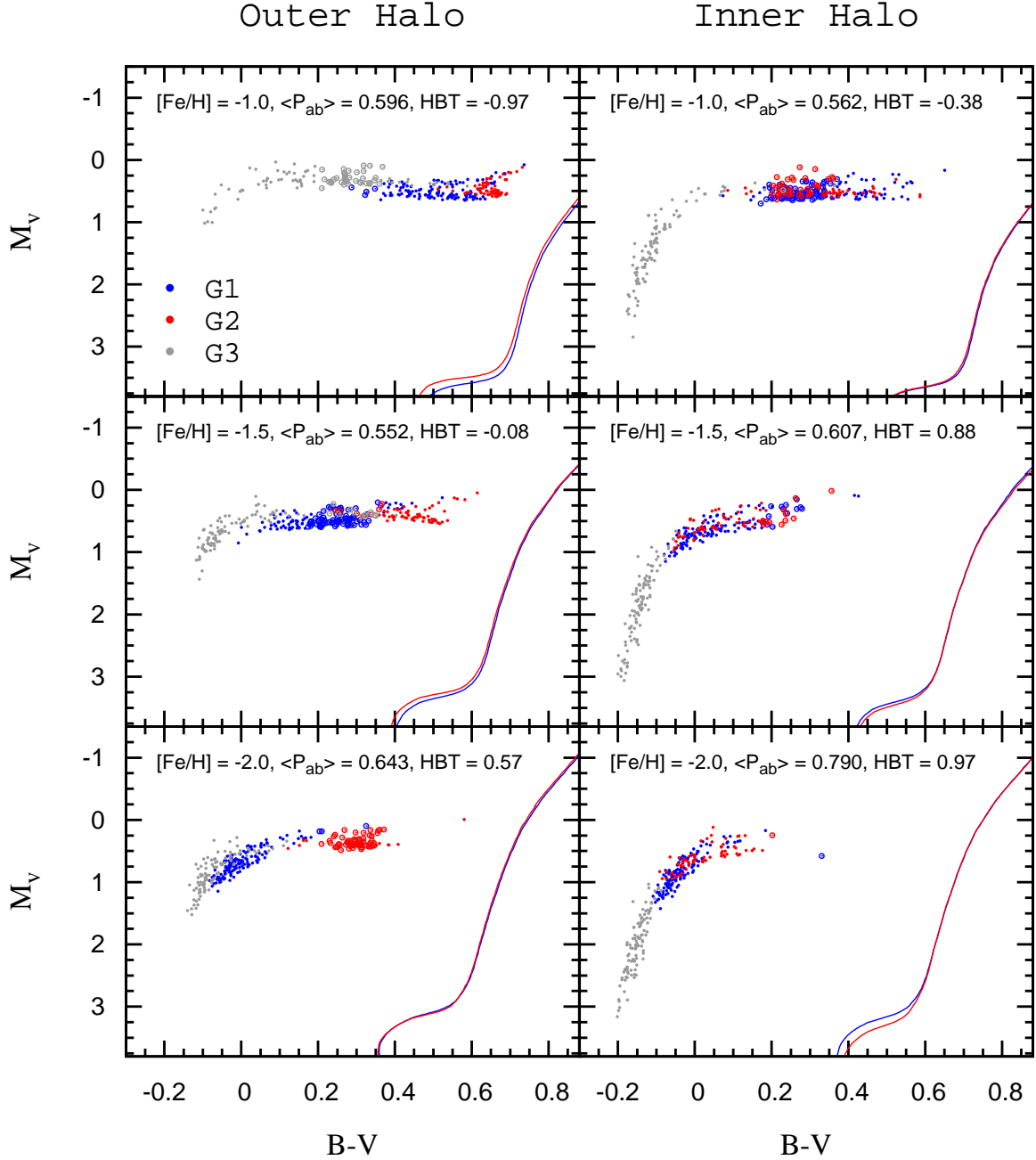}
\caption{Selection of our synthetic HB models used in the construction of model lines in Figure~\ref{fig6}. $\left<P_{\rm ab}\right>$ and HB type values are from G1 and G2. For GCs with G3, HB stars belonging to this additional subpopulation are presented as filled gray circles. Open circles are RR Lyrae variables and $\left<P_{\rm ab}\right>$ is in days.\label{fig7}}
\end{figure}

\begin{figure}
\epsscale{1.05}
\plotone{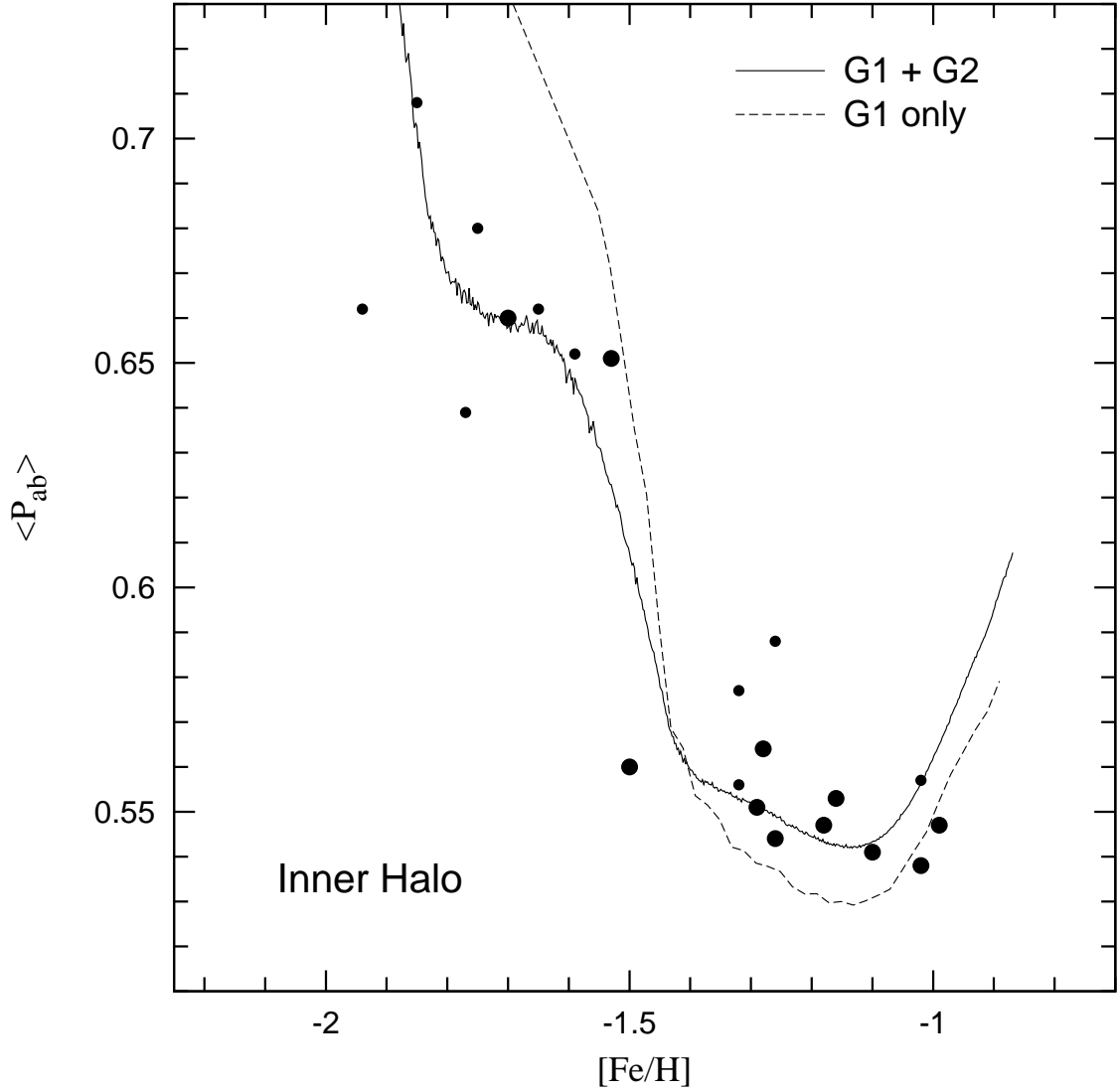}
\caption{The ``two-stage jump" in $\left<P_{\rm ab}\right>$ in the inner halo GCs. The dashed and solid lines are our models constructed with only G1 and with G1+G2, respectively. Note that the observed dichotomy can be reproduced only in the multiple population paradigm (i.e. G1+G2 model, see the text). \label{fig8}}
\end{figure}

There are several remaining issues to be clarified related to this work. For example, RR Lyrae variables in the Local Group dwarf galaxies often show Oosterhoff intermediate characteristics with $\left<P_{\rm ab}\right>$ of $\sim$0.60 day \citep[see, e.g.,][]{Cat09}. Although Figures~\ref{fig3} and \ref{fig4} indicate that some of our models (see the dashed lines), with different assumptions on the ages of G1 and G2, can be placed within the Oosterhoff gap, further investigations are required to fully understand this phenomenon. As to the observational test, as emphasized by \citet{Jan14}, spectroscopic observations for the Na, He, and CNO abundances can provide important tests on our placements of G1, G2, and G3 on the HB. Although the available observations \citep{Gra12b,Gra13,Gra14,Vil12,Mar13,Mar14} are not in conflict with our models \citep[see][]{Jan14}, especially because now our models (see Figure~\ref{fig7}) show more flexibility on the placements of G1 and G2 depending on $\Delta$t(G1 - G2), spectroscopy for a large sample of RR Lyrae and HB stars would certainly help to better understand the population-shift scenario suggested in our models. If confirmed by further observations and models, our understanding for this long standing problem, the Oosterhoff dichotomy, is now sufficiently well-advanced that the RR Lyrae stars can be used as more reliable standard candle and galaxy formation tracers.




\acknowledgments
We thank the referee for a number of helpful suggestions. We also thank Hansung Gim for his early contribution in HB morphology classification, and Seok-Joo Joo for his FORTRAN code, based on which the present code was developed to include multiple populations and the details of RR Lyrae stars. Support for this work was provided by the National Research Foundation of Korea to the Center for Galaxy Evolution Research.

\clearpage



\clearpage

\clearpage

\clearpage

\clearpage

\clearpage

\clearpage

\clearpage

\begin{figure}
\epsscale{.85}
\plotone{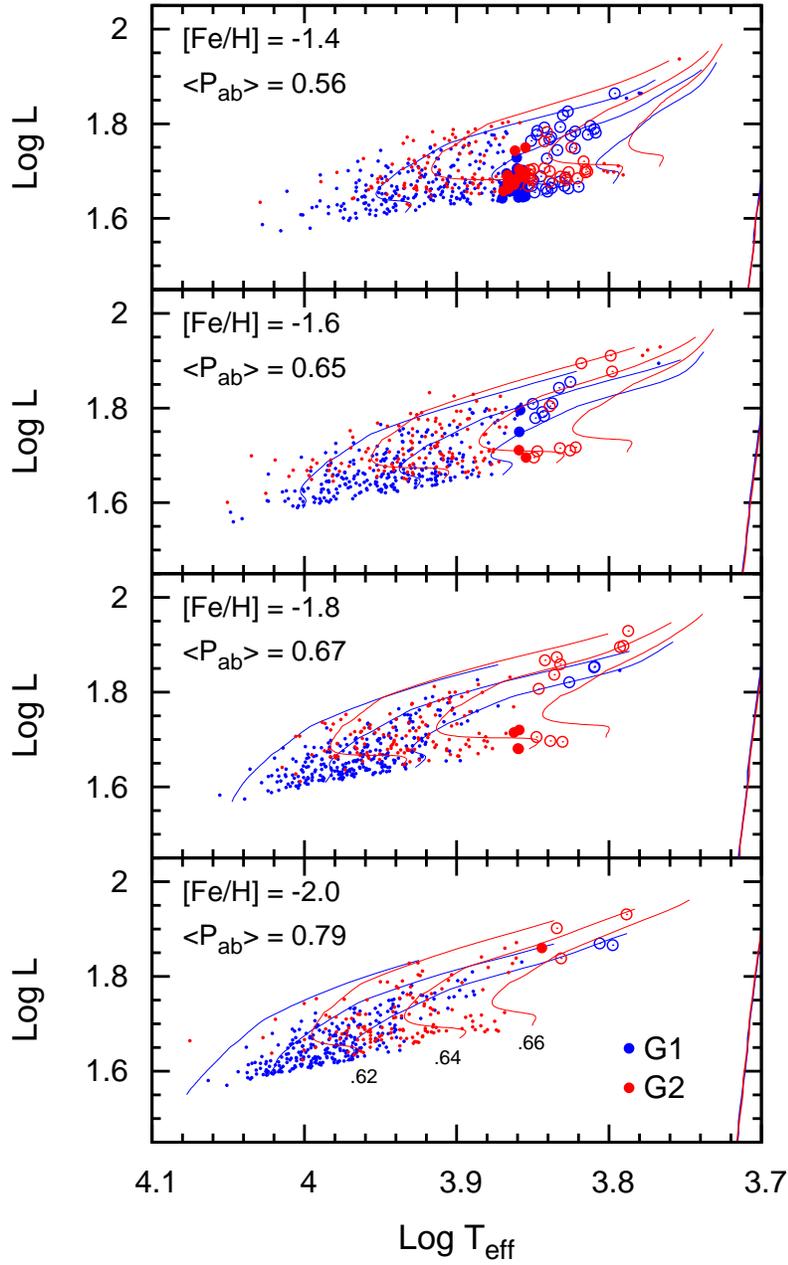}
\caption{Selection of our synthetic HB models (with G1 and G2) for the inner halo GCs, which explain the ``population-shift" within the IS and the consequent ``two-stage jump" in $\left<P_{\rm ab}\right>$.  In each panel, [Fe/H] and $\left<P_{\rm ab}\right>$ in days are denoted, which correspond to the solid line in Figure~\ref{fig8}. Large open (filled) circles are type ab (type c) RR Lyrae variables. The HB evolutionary tracks are overlaid (shown by the solid lines) and labeled by its total mass in solar units. \label{fig9}}
\end{figure}

\clearpage

\begin{figure}
\epsscale{0.800}
\plotone{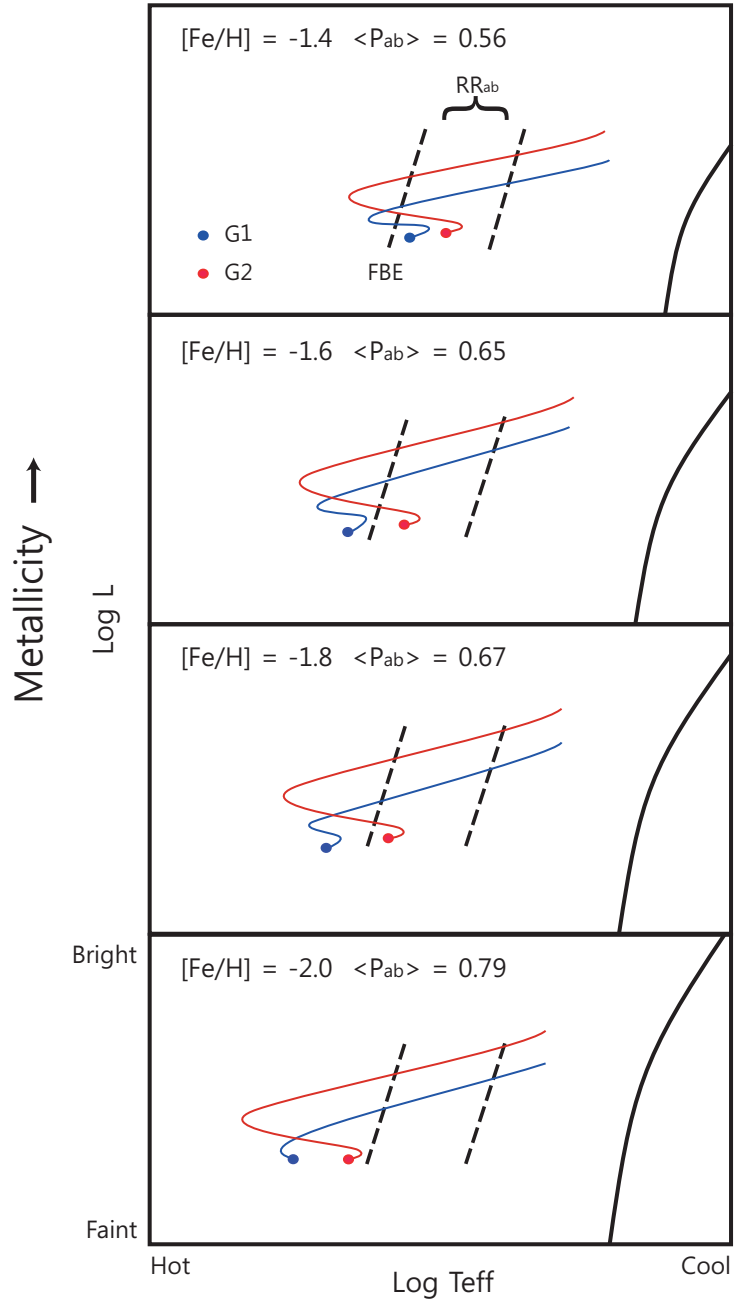}
\caption{Schematic diagram illustrating the population-shift within the type ab RR Lyrae zone (thick dashed lines) for the inner halo GCs. Note that the ZAHB portion of the evolutionary track departs sequentially from G1 to G2 with decreasing metallicity, which leads to the two-stage jump in $\left<P_{\rm ab}\right>$ (see the text).  \label{fig10}}
\end{figure}

\clearpage

\begin{figure}
\epsscale{1.05}
\plotone{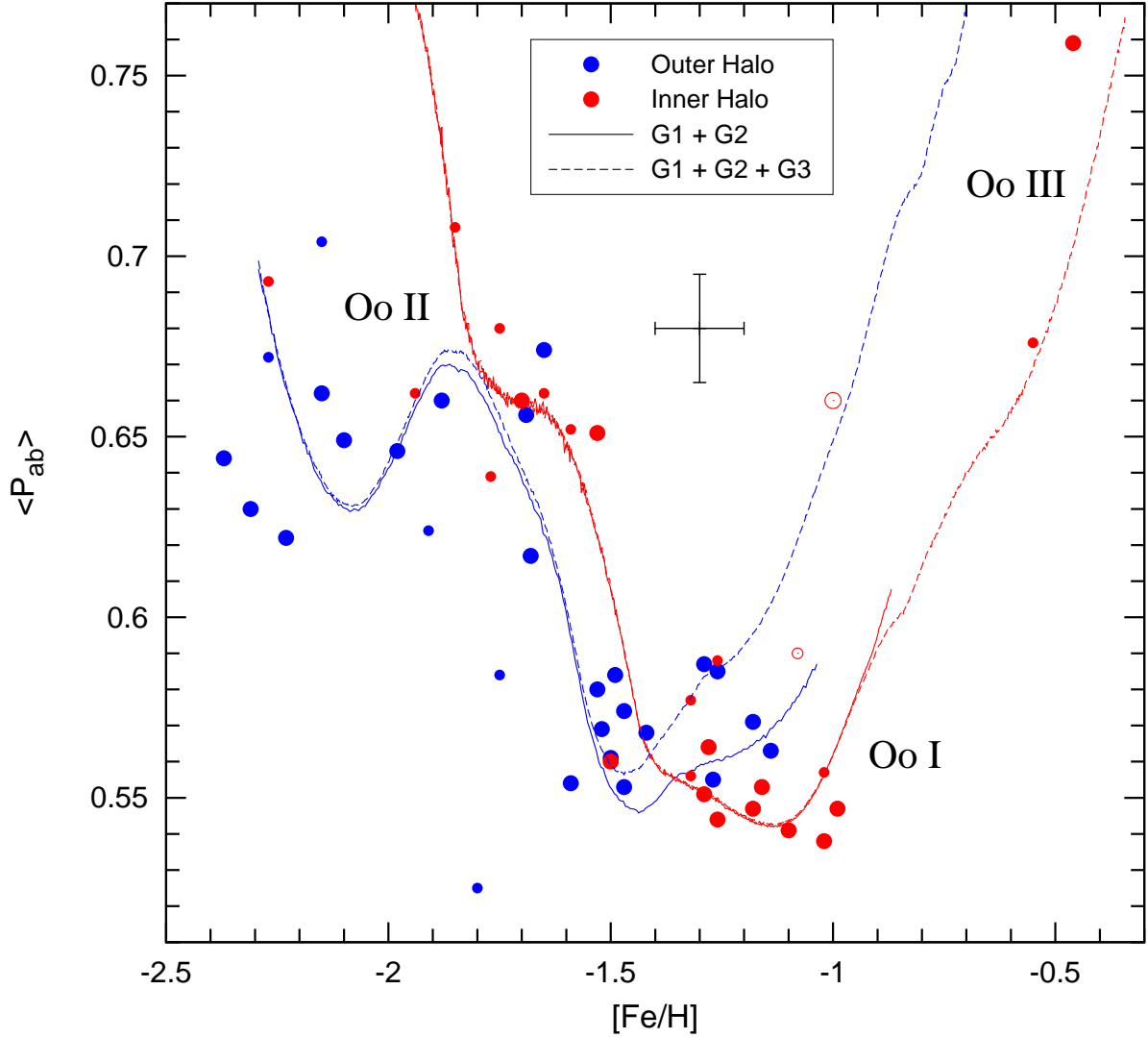}
\caption{Same as Figure~\ref{fig1}, but with our best-fit models including G3. Note that four bulge GCs (NGC 6441, 6388, 2Mass-GC02, and Terzan 10), and some group I GCs in the outer halo are better reproduced by our models with G3 (the dashed lines). \label{fig11}}
\end{figure}

\clearpage

\begin{figure}
\epsscale{1.05}
\plotone{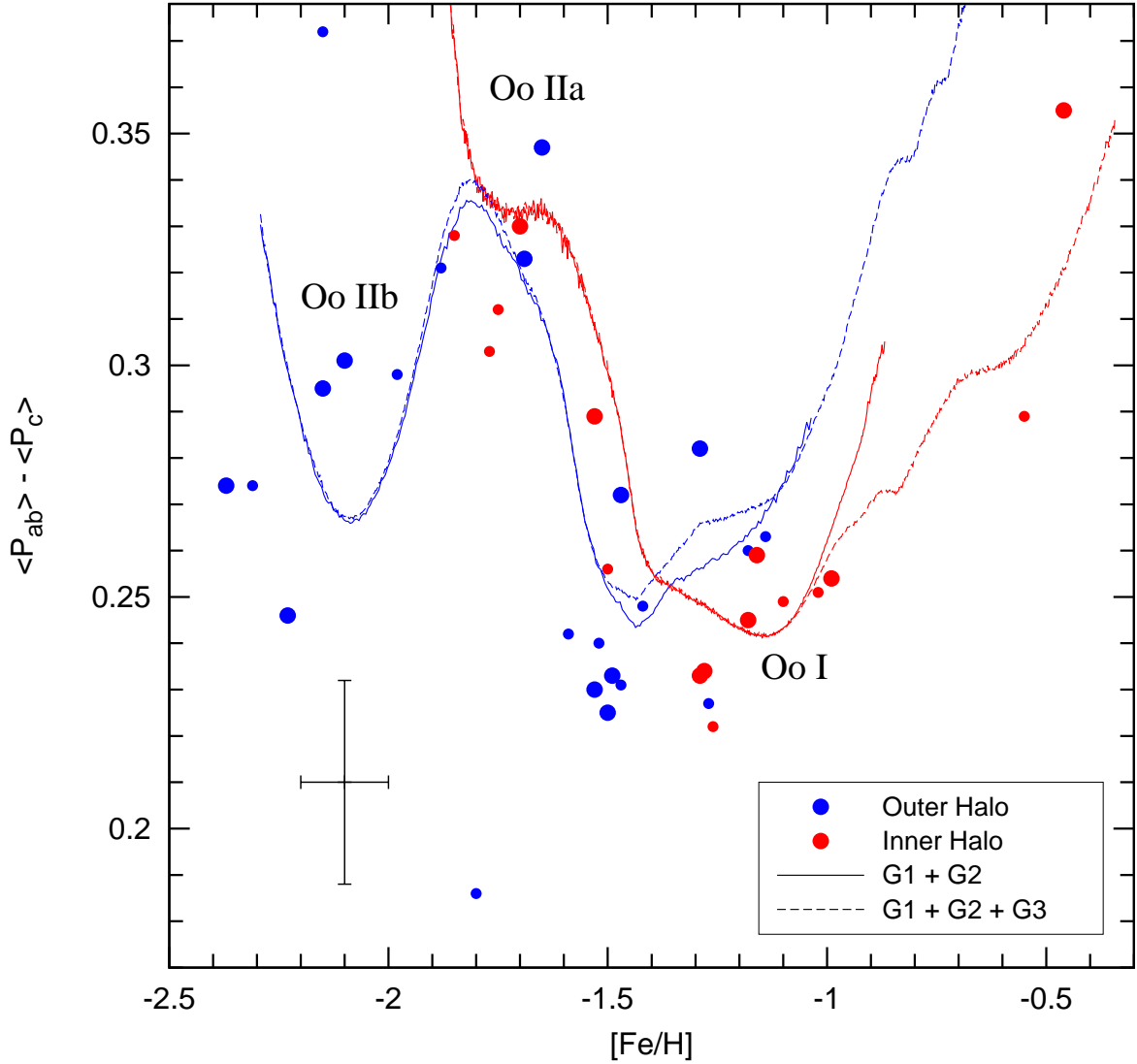}
\caption{Similar to Figure~\ref{fig11}, but for the difference in the mean periods between type ab and type c variables, $\left<P_{\rm ab}\right>$ - $\left<P_{\rm c}\right>$. The large (small) symbols are for clusters with $N$ $\geqq$ 10 (3 $\leqq$ $N$ $<$ 10) for both ab and c type variables, respectively. The typical error bar is shown (see the text).  \label{fig12}}
\end{figure}

\clearpage







\clearpage

\begin{deluxetable}{lccccccccc}
\tablecolumns{10}
\setlength{\tabcolsep}{7pt}
\tablecaption{Milky Way Globular Clusters with three or more type ab RR Lyrae Variables\label{tab1}}

\tablehead{
ID & Name & [Fe/H] & $<P_{ab}>$ & $N_{ab}$ & $<P_{c}>$ & $N_{c}$ & HB type & $R_{gc}$ & \colhead{}{}{Source} 
}
\startdata
NGC 362 & & -1.26&  0.585&  24&  0.365&     3&   -0.82&     9.4&    1 \\
NGC 1261& & -1.27&  0.555& 13&  0.328&     5&   -0.71&    18.1&  \\
NGC 1851&&    -1.18&    0.571&    21&    0.311&     8&       -0.31&     16.6& \\
NGC 2419&&   -2.15&    0.662&    38&    0.367&     37&      0.76&      89.9&    2 \\
NGC 2808&&        -1.14&    0.563&    11&    0.300&     5&       0.09&      11.1   & 3 \\
NGC 3201&&    -1.59&    0.554&    72&    0.312&     5&       0.08&      8.8&     \\
NGC 4147&&        -1.80&    0.525&    5&     0.339&     10&      0.55&      21.4&    4 \\
NGC 4590&   M68&  -2.23&    0.622&    14&    0.376&     32&      0.44&      10.2&    5 \\
NGC 4833&&        -1.85&    0.708&    7&     0.380&     7&       0.95&      7.0 \\
NGC 5024&   M53&  -2.10&    0.649&    29&    0.348&     29&      0.81&      18.4 \\
NGC 5053&&        -2.27&    0.672&    5&     0.354&     4&      0.54&      17.8  \\
NGC 5139&  $\omega$ Cen& -1.53&    0.651&    76&    0.362&     85&      0.90&      6.4 \\
NGC 5272&   M3&   -1.50&    0.561&    158&   0.336&     48&      0.08&      12.0&    6 \\
NGC 5286&&        -1.69&    0.656&    30&    0.333&     22&      0.74&      8.9&     7\\
NGC 5466&&        -1.98&    0.646&    13&    0.348&     7&       0.59&      16.3 \\
NGC 5634&&        -1.88&    0.660&     12&    0.339&     7&       0.94&      21.2&    8\\
NGC 5824&&        -1.91&    0.624&    7&     \nodata &    0&       0.91&      25.9 \\
NGC 5904&   M5&   -1.29&    0.551&    91&    0.318&     35&      0.39&      6.2 \\
NGC 5986&&        -1.59&    0.652&    7&     0.328&     1&       0.94&      4.8 \\
NGC 6093&   M80&  -1.75&    0.680&    7&     0.368&     7&       0.93&      3.8&    9 \\
NGC 6121&   M4&   -1.16&    0.553&    31&    0.294&     13&      -0.07&     5.9&     10\\
NGC 6139&&        -1.65&    0.662&    3&     0.417&     1&        0.97&      3.6 \\
NGC 6171&   M107& -1.02&    0.538&    15&    0.287&     7&       -0.76&     3.3 \\

NGC 6229&&        -1.47&    0.553&    30&    0.322&     8&       0.36&      29.8 \\
NGC 6266&   M62&  -1.18&    0.547&    133&   0.302&     76&      0.60&      1.7&     11\\
NGC 6284&&        -1.26&    0.588&    6&     \nodata&        0&       0.90&     7.5 \\ 
NGC 6333&   M9&   -1.77&    0.639&    8&     0.336&     10&      0.85&      1.7&     12\\
NGC 6341&   M92&  -2.31&    0.630&     11&    0.356&     6&       0.87&     9.6   \\
NGC 6362&&        -0.99&    0.547&    18&    0.293&     17&      -0.45&     5.1   \\
NGC 6388&&        -0.55&    0.676&    9&     0.387&     11&      -0.75&     3.1&     13\\
NGC 6402&   M14&  -1.28&    0.564&    39&    0.330&     15&      0.82&      4.0   \\
NGC 6426&&        -2.15&    0.704&    9&     0.332&     4&       0.53&      14.4  \\
NGC 6441&&        -0.46&    0.759&    42&    0.404&     25&      -0.81&     3.9&   13 \\
NGC 6558&&        -1.32&    0.556&    6&     0.345&     3&       0.84&      1.0 \\
NGC 6584&&        -1.50&    0.560&     34&    0.304&     8&       -0.09&    7.0 \\
NGC 6626&   M28&  -1.32&    0.577&    8&     0.312&     2&       0.96&      2.7 \\
NGC 6642&&        -1.26&    0.544&    10&    0.322&     6&       0.43&      1.7 \\
NGC 6656&   M22&  -1.70&    0.660&    10&    0.330&     16&      0.97&      4.9&     14 \\
NGC 6712&&        -1.02&    0.557&    7&     0.338&     2&       -0.66&     3.5 \\
NGC 6715&   M54&  -1.49&    0.584&    95&    0.351&     33&      0.75&      18.9&    15 \\
NGC 6723&&        -1.10&    0.541&    35&    0.292&     7&       0.00&      2.6&     16\\
NGC 6809&   M55&  -1.94&    0.662&    4&     0.340&     9&       0.88&      3.9  \\  
NGC 6864&   M75&  -1.29&    0.587&    25&    0.305&     13&      -0.14&     14.7&    17 \\
NGC 6934&&        -1.47&    0.574&    68&    0.302&     10&      0.21&      12.8 \\
NGC 6981&   M72&  -1.42&    0.568&    36&    0.320&     7&       0.07&      12.9&    18 \\
NGC 7006&&        -1.52&    0.569&    53&    0.329&     9&       -0.11&     38.5\\
NGC 7078&   M15&  -2.37&    0.644&    68&    0.370&     95&      0.74&      10.4&    19\\
NGC 7089&   M2&   -1.65&    0.674&    19&    0.327&     13&      0.92&      10.4&    20 \\

NGC 7099&   M30&  -2.27&    0.693&    4&    \nodata  &  0&     0.94&      7.1&   21 \\
Arp2&&            -1.75&    0.584&    8&     0.292&     1&       0.53&      21.4&    8\\
IC 4499&&         -1.53&    0.580&    63&    0.350&     34&      0.12&      15.7 \\
Rup 106&&          -1.68&    0.617&    13&    \nodata&       0&       -0.82&     18.5 \\
Terzan 10 && -1.00&0.660&10&\nodata&\nodata&\nodata& 2.3&22 \\
2Mass-GC2 &&-1.08&0.590&3&\nodata&\nodata&\nodata&3.2&22 \\

\enddata


\tablerefs{
(1) \citealt{Sze07}; (2) \citealt{Dic11}[ (3) \citealt{Kun13c};
(4) \citealt{Ste05}; (5) \citealt{Sar14}; (6) \citealt{Cor01};
(7) \citealt{Zor10}; (8) \citealt{Sal05}; (9) \citealt{Kop13};
(10) \citealt{Ste14}; (11) \citealt{Con10}; (12) \citealt{Are13};
(13) \citealt{Cor06}; (14) \citealt{Kun13b}; (15) \citealt{Sol10}; (16) \citealt{JW14};
(17) \citealt{Cor03}; (18) \citealt{Ami13}; (19) \citealt{Cor08}; (20) \citealt{Laz06};
(21) \citealt{Kai13}; (22) \citealt{Alo15}.
}
\end{deluxetable}

\clearpage

\begin{deluxetable}{cccccccc}
\tablecolumns{8}
\tablewidth{0pc}
\tablecaption{Parameters from Our Best-fit Simulations\label{tab2}}
\tablehead{

\colhead{}    &  \multicolumn{3}{c}{Outer halo GCs} &   \colhead{}   &
\multicolumn{3}{c}{Inner halo GCs} \\

\cline{2-4} \cline{6-8} \\
\colhead{Population\tablenotemark{a}} & \colhead{Age (Gyr)}   & \colhead{{\it Y}}    & \colhead{$\Delta Z_{CNO}$} &
\colhead{}    & \colhead{Age (Gyr)}   & \colhead{{\it Y}}    & \colhead{$\Delta Z_{CNO}$}}
\startdata
G1 & 12.4 & 0.23 & 0 &
& 13.2 & 0.23 & 0 \\
G2 & 11.0 & 0.24 & 0.00035 &
& 12.7 & 0.24 & 0.00035 \\
G3 & 10.5-10.3 & 0.27-0.30 & 0.00035 &
& 12.2-12.0 & 0.27-0.30 & 0.00035 \\

\enddata

\tablenotetext{a}{Papulation ratio G1 : G2 : G3 = 0.37 : 0.20 : 0.43 }
\end{deluxetable}

\clearpage

\begin{deluxetable}{cccccccccccc}
\tablecolumns{12}
\setlength{\tabcolsep}{3pt}

\tablecaption{Results of Our Best-fit Models\tablenotemark{a}\label{tab3}}
\tablehead{
\colhead{}    &  \multicolumn{5}{c}{Outer halo GCs} &   \colhead{}   &
\multicolumn{5}{c}{Inner halo GCs} \\
\cline{2-6} \cline{8-12} \\

\colhead{} & \multicolumn{2}{c}{G1+G2} & \colhead{} &
\multicolumn{2}{c}{G1+G2+G3} & \colhead{} & \multicolumn{2}{c}{G1+G2} & \colhead{} &
\multicolumn{2}{c}{G1+G2+G3} \\
\cline{2-3} \cline{5-6} \cline{8-9} \cline{11-12}
\colhead{[Fe/H]} & \colhead{$<P_{ab}>$}   & \colhead{HB type}    & \colhead{} & \colhead{$<P_{ab}>$} &
\colhead{HB type}    & \colhead{} & \colhead{$<P_{ab}>$}   & \colhead{HB type}    & \colhead{} & 
\colhead{$<P_{ab}>$} & \colhead{HB type}}
\startdata

-2.25 & 0.677 &  0.72 && 0.677 &  0.84 && \nodata&1.00  && \nodata& 1.00 \\
-2.20 & 0.653 &  0.69 && 0.653 &  0.82 && 0.849 &  0.99 &&  \nodata&  1.00\\
-2.15 & 0.638 &  0.65 && 0.639 &  0.80 && 0.833 &  0.99 &&  0.833&  0.99\\
-2.10 & 0.630 &  0.62 && 0.632 &  0.78 && 0.819 &  0.99 &&  0.821&  0.99\\
-2.05 & 0.631 &  0.60 && 0.633 &  0.77 && 0.806 &  0.99 &&  0.808&  0.99\\
-2.00 & 0.641 &  0.56 && 0.642 &  0.75 && 0.794 &  0.98 &&  0.798&  0.99\\
-1.95 & 0.654 &  0.51 && 0.656 &  0.72 && 0.775 &  0.97 &&  0.772&  0.98\\
-1.90 & 0.667 &  0.43 && 0.669 &  0.67 && 0.749 &  0.96 &&  0.748&  0.98\\
-1.85 & 0.669 &  0.34 && 0.674 &  0.62 && 0.704 &  0.95 &&  0.705&  0.97\\
-1.80 & 0.664 &  0.29 && 0.669 &  0.59 && 0.672 &  0.95 &&  0.672&  0.97\\
-1.75 & 0.652 &  0.26 && 0.656 &  0.57 && 0.662 &  0.95 &&  0.664&  0.97\\
-1.70 & 0.639 &  0.23 && 0.642 &  0.55 && 0.659 &  0.95 &&  0.658&  0.97\\
-1.65 & 0.625 &  0.20 && 0.627 &  0.53 && 0.656 &  0.95 &&  0.658&  0.97\\
-1.60 & 0.602 &  0.13 && 0.605 &  0.49 && 0.648 &  0.94 &&  0.649&  0.96\\
-1.55 & 0.571 &  0.03 && 0.574 &  0.42 && 0.632 &  0.92 &&  0.632&  0.95\\
-1.50 & 0.553 & -0.09 && 0.560 &  0.33 && 0.609 &  0.89 &&  0.609&  0.93\\
-1.45 & 0.547 & -0.21 && 0.558 &  0.24 && 0.580 &  0.83 &&  0.581&  0.90\\
-1.40 & 0.549 & -0.35 && 0.564 &  0.14 && 0.560 &  0.77 &&  0.560&  0.87\\
-1.35 & 0.556 & -0.50 && 0.572 &  0.02 && 0.555 &  0.69 &&  0.555&  0.82\\
-1.30 & 0.559 & -0.64 && 0.581 & -0.09 && 0.553 &  0.58 &&  0.552&  0.76\\
-1.25 & 0.561 & -0.70 && 0.586 & -0.14 && 0.548 &  0.50 &&  0.549&  0.71\\
-1.20 & 0.563 & -0.75 && 0.592 & -0.20 && 0.545 &  0.40 &&  0.545&  0.65\\
-1.15 & 0.568 & -0.82 && 0.601 & -0.26 && 0.542 &  0.26 &&  0.543&  0.57\\
-1.10 & 0.574 & -0.88 && 0.614 & -0.33 && 0.543 &  0.06 &&  0.544&  0.46\\
-1.05 & 0.584 & -0.94 && 0.631 & -0.42 && 0.549 & -0.15 &&  0.550&  0.33\\
-1.00 &  \nodata     &    \nodata    && 0.648 & -0.51 && 0.562 & -0.38 &&  0.562&  0.18\\
-0.95 &   \nodata     &   \nodata     && 0.667 & -0.59 && 0.578 & -0.63 &&  0.577&  0.02\\
-0.90 &   \nodata    &   \nodata     && 0.691 & -0.68 && 0.595 & -0.86 &&  0.591& -0.14\\
-0.85 &   \nodata     &  \nodata      && 0.713 & -0.76 &&   \nodata     &  \nodata      &&  0.600& -0.25\\
-0.80 & \nodata       &    \nodata    && 0.723 & -0.83 &&   \nodata     &  \nodata     &&  0.614& -0.32\\
-0.75 &  \nodata      &    \nodata    && 0.747 & -0.89 &&   \nodata     & \nodata      &&  0.627& -0.39\\
-0.70 &   \nodata     & \nodata     && 0.769 & -0.93 &&    \nodata    &  \nodata      &&  0.640& -0.45\\
-0.65 &  \nodata      &   \nodata     &&   \nodata     &  \nodata      &&   \nodata    & \nodata      &&  0.649& -0.52\\
-0.60 &   \nodata     &  \nodata      &&  \nodata     &    \nodata   &&  \nodata      &  \nodata     &&  0.656& -0.59\\
-0.55 &  \nodata      & \nodata      &&   \nodata     &   \nodata     &&\nodata     & \nodata     &&  0.669& -0.66\\
-0.50 &  \nodata      &   \nodata     &&  \nodata     &   \nodata     && \nodata       & \nodata      &&  0.688& -0.74\\
-0.45 & \nodata       &  \nodata      && \nodata       &  \nodata      &&   \nodata     & \nodata      &&  0.708& -0.81\\
-0.40 &   \nodata     &  \nodata    &&  \nodata     &    \nodata    &&  \nodata     &\nodata       &&  0.734& -0.88\\
-0.35 &  \nodata      &  \nodata      && \nodata       &   \nodata     &&  \nodata      &  \nodata      &&  0.761& -0.94\\

\enddata

\tablenotetext{a}{$\left<P_{\rm ab}\right>$ is in days}
\end{deluxetable}

\clearpage




\end{document}